\documentclass[12pt]{iopart}
\usepackage{amssymb}
\expandafter\let\csname equation*\endcsname\relax
\expandafter\let\csname endequation*\endcsname\relax

\usepackage{amsmath}

\usepackage[longnamesfirst, square,sort,comma,numbers]{natbib}
\usepackage{graphicx}
\usepackage{dcolumn}
\usepackage{bm}
\usepackage[mathlines,running]{lineno}
\usepackage{afterpage}
\usepackage{etoolbox}
\usepackage{subcaption}
\usepackage{url} 
\usepackage[dvipsnames]{xcolor}
\usepackage{tikz}
\usetikzlibrary{arrows}
\usetikzlibrary{shapes}
\usetikzlibrary{patterns}
\usetikzlibrary{calc,math}
\usetikzlibrary{calc,math}

\usepackage{float}

\usepackage{diagbox, xcolor}
\newcommand{\tL}{\tilde{L}}
\newcommand{\bfx}{\mathbf{x}}

\newcommand{\bfX}{\mathbf{X}}

\begin{document}

\title{Quantitative Assessment of PINN Inference on Experimental Data for Gravity Currents Flows}{Quantitative Assessment of PINN Inference on Experimental Data for Gravity Currents Flows}

\author{Mickaël Delcey$^1$, Yoann Cheny$^1$, Jean Schneider$^1$, Simon Becker$^1$, Sébastien Kiesgen De Richter$^{1,2}$}
\address{$^1$LEMTA, Université de Lorraine, CNRS, 2, Avenue de la Forêt de Haye, B.P. 160, 54500, Vandœuvre-lès-Nancy}
\address{$^2$Institut Universitaire de France (IUF)}
\ead{mickael.delcey@loria.fr}
\vspace{5pt}

\begin{indented}
\item[]August 2017
\end{indented}

\begin{abstract}
In this paper, we apply Physics Informed Neural Networks (PINNs) to infer velocity and pressure field from Light Attenuation Technique (LAT) measurements for gravity current induced by lock-exchange. In a PINN model, physical laws are embedded in the loss function of a neural network, such that the model fits the training data but is also constrained to reduce the residuals of the governing equations. PINNs are able to solve ill-posed inverse problems training on sparse and noisy data, and therefore can be applied to real engineering applications. 
The noise robustness of PINNs and the model parameters are investigated in a 2 dimensions toy case on a lock-exchange configuration, employing synthetic data. Then  we train a PINN with experimental LAT measurements and quantitatively compare the velocity fields inferred to Particle Image Velocimetry (PIV) measurements performed simultaneously on the same experiment.The results state that accurate and useful quantities can be derived from a PINN model trained on real experimental data which is encouraging for a better description of gravity currents.
\end{abstract}


\section*{Introduction}

In recent years, advances in deep learning \cite{Goodfellow} and the universal approximation nature of neural networks offer new possibilities in fluid mechanics \cite{brunton2020machine}. However, some physical systems remain beyond the reach of purely data-driven models due to their chaotic nature and limited data availability. 
To address this, Raissi et al. \cite{raissi2020science} recently introcuded the Physics-Informed Neural Network (PINNs) approach. PINNs can solve forward and inverse partial differential equation (PDEs) problems by incorporating prior knowledge of physical laws, constraining the neural network to adhere to a set of PDEs. This approach has quickly demonstrated its effectiveness on various non-linear problems, both forward \cite{mao2020physics}, and inverse \cite{jagtap2022physics} (see \cite{cuomo2022scientific} for a complete review). Physics-Informed Neural Networks (PINNs) are are particularly suitable for dealing with noisy and incomplete data, as well as ill-posed problems, making them ideal for addressing complex reconstruction problems in fluid dynamics. In fluid mechanics, reconstruction problems aim to determine the complete flow field, denoted as $\boldsymbol{v}$, from limited information, such as partial measurements or incomplete physics. One of the most challenging types of reconstruction problems is inferring the full velocity and pressure fields from partial measurements of a single component of $\boldsymbol{v}$, such as concentration, temperature, or density fields, with the constraint that only one realization of the flow can be performed. This is particularly difficult when dealing with complex, non-linear governing equations like the Navier-Stokes equations. In this context, linear methods such as Proper Orthogonal Decomposition (POD) \cite{berkooz1993proper}, Dynamic Mode Decomposition (DMD) \cite{schmid2010dynamic}, and standard regression methods  are inadequate due to the inherent nonlinearity of the problem and the lack of sufficient data on the quantities of interest. These methods fail to capture the complex interactions within the flow fields accurately. Additionally, data assimilation techniques \cite{rodell2004global}, while powerful, are not suitable due to the ill-posed nature of the problem and their computational intensity. Given these challenges, PINNs emerge as one of the few viable methods that can effectively address this reconstruction problem.
In \cite{raissi2018hidden}, Raissi et al. successfully reconstructed the velocity and pressure fields from synthetic measurements of passive scalar concentration for the canonical flow past a cylinder. This method was also tested with Gaussian noise, demonstrating its robustness under noisy conditions. In  \cite{cai2021physics} the authors reconstructed accurately the velocity and pressure fields from temperature field in a synthetical case of forced convection. While PINNs have been frequently used for reconstruction problems on synthetic data, there are few examples of their application to experimental data. One major contribution is \cite{esspresso2021flow}, in which Kaniadakis et al. used a PINNs model to infer the velocity and pressure fields from experimental measurements of temperature obtained via Background Oriented Schlieren (BOS) \cite{BOS}  for a flow over an espresso cup. However, the velocity fields inferred were only validated through a qualitative comparison with Particle Image Velocimetry (PIV) \cite{perez2018piv} measurements in a companion experiment. Recently, PINNs have also been applied to BOS data to reconstruct high-speed flow fields \cite{exp1}, but no validation of the velocity field is available.
In this paper, we aim to take a step further and demonstrate a quantitative comparison between the velocity fields predicted by PINNs and those measured by PIV. To achieve this, we consider gravity currents \cite{ungarish2009introduction} the canonical flow of this physical phenomenon: the lock exchange \cite{shin}, which we have previously studied in another work \cite{pof}. This flow is complex and governed by the Navier-Stokes equations coupled with the transport of density. Our experimental setup couples density measurements obtained by the Light Attenuation Technique (LAT) \cite{dossmann2016experiments} and velocity measurements by PIV in the same experiment. We will train a PINN model using the LAT density data and compare the velocity fields predicted by the model with those obtained from PIV measurements for a single realization of the flow. In all subsequent discussions, we will consider only the 2D governing equations. Conducting such a comparison requires making a strong assumption of a 2D model, which is commonly made in the study of gravity currents \cite{inbook}. This assumption will be discussed at the end of the article.
The governing equations of gravity currents flow and the principles of Physics informed Neural Networks will be discussed in section \ref{section1} and \ref{section2}, respectively.
In section \ref{section3} we calibrate the model hyperparameters on a digital twin of the experimental apparatus of the lock exchange configuration and study the noise robustness of PINNs on synthetical data.
In Section \ref{section4}, we will focus on the application of PINNs to experimental data from the lock-exchange experiment. We will use the PINNs model to reconstruct the velocity and pressure fields from LAT experimental measurements and then compare these reconstructed velocity fields with PIV measurements obtained simultaneously during the same experiment.
Finally, we will discuss the potential outcomes of the trained model for gravity currents.

\section{Governing equations of Lock-Exchange gravity currents}\label{section1}

\begin{figure}[H]
\centering
\includegraphics[scale= 0.2]{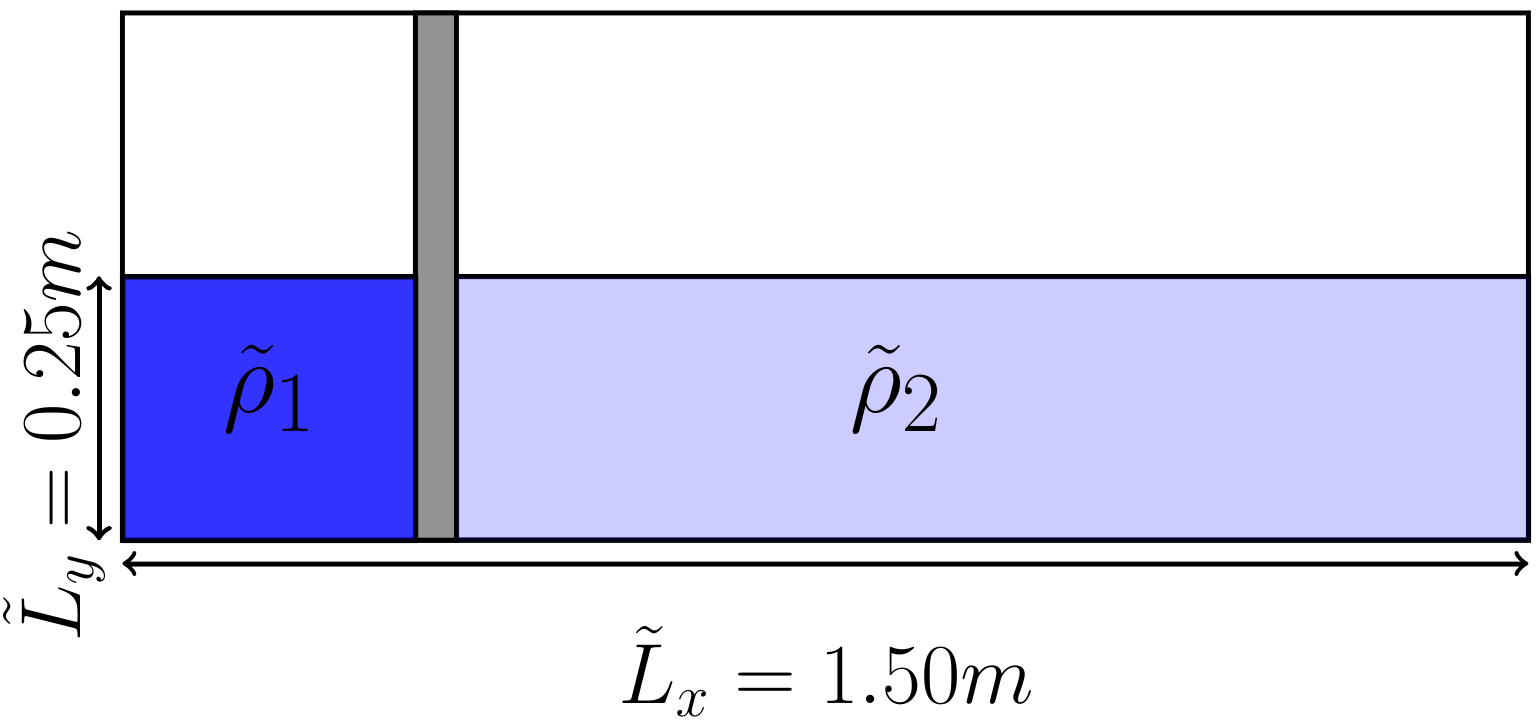}
\caption{Lock exchange tank : The lock-exchange configuration. At the initial moment, the wall is removed and the system evolves under the action of gravity only. The difference of density is achieved by a difference of salinity.}    \label{fig:cuve}
\end{figure}

In the lock exchange flow \cite{shin}, fluids of different densities initially at
rest are separated by a vertical membrane, in a tank. The fluids in the left and right hand compartments have the respective densities $\tilde{\rho_1}$ and $\tilde{\rho_2}$ with  $\tilde{\rho_1} > \tilde{\rho_2}$, the density gradient
is caused for instance by salinity difference. Once the membrane is withdrawn, the density difference $\Delta\rho = \tilde{\rho_1} - \tilde{\rho_2}$ between the two compartments induces the generation and propagation of a gravity current at the tank bottom.
The intensity of turbulent structures and induced mixing in the current are controlled by the Reynolds number ($Re$), which is defined as $Re = \frac{UL}{\nu}$, where $U$ is the characteristic velocity, $L$ is the characteristic length, and $\nu$ is the kinematic viscosity of the fluid. Additionally, the mixing of scalar quantities is characterized by the Schmidt number ($Sc$), defined as $Sc = \frac{\nu}{D}$, where $D$ is the mass diffusivity. The flow is governed by Navier-Stokes, mass transport and incompressibility equations. Moreover, in this paper, we study gravity currents induced by small difference in density, so that Boussinesq Aproximation \cite{gray1976validity} is  effective.  A sketch of the tank setup can be seen in Fig.~\ref{fig:cuve}.

\definecolor{op}{gray}{0.9}

In order to render the equations dimensionless (see ~\cite{hartel2000analysis}), we use $\tL_y$ as the characteristic
length scale and the characteristic velocity is the buoyancy velocity defined as~:

\begin{equation}\label{eq:buyvel}
    \tilde{u}_b=\sqrt{g\frac{\tilde{\rho_1}-\tilde{\rho_2}}{\tilde{\rho_a}}\tilde{L}_y}, \quad \text{with}\quad\tilde{\rho}_a=\frac{\tilde{\rho_1}+\tilde{\rho_2}}{2}.
\end{equation}
We consider small density difference $\tilde{\rho}_1 - \tilde{\rho}_2\ll \tilde{\rho}_1$ for which the Boussinesq approximation can be adopted, thus the governing equations read in dimensionless form~:


\begin{subequations}
\label{eq:NS}
\begin{align}
    & \dfrac{\partial \boldsymbol{u}}{\partial t} + (\boldsymbol{u} \cdot \nabla )~ \boldsymbol{u}  +\nabla p - \dfrac{1}{Re}\nabla ^{2} \boldsymbol{u}  \label{eq:momx} + \rho ~ \boldsymbol{e_{y}}=0,\\
    & \dfrac{\partial \rho}{\partial t}+(\boldsymbol{ u} \cdot \nabla )~ \rho - \dfrac{1}{Re\,Sc}\nabla^{2}\rho \label{eq:mass}=0,\\
    & \nabla \cdot \boldsymbol{ u} = 0\label{eq:cont},  
\end{align}
\end{subequations}

where  $\boldsymbol{u} = (u,v)^\intercal$ denotes the dimensionless velocity vector. The non-dimensional pressure $p$ and density $\rho$ are given by:
\begin{equation}\label{eq:prhoadim}
    p=\frac{\tilde{p}}{\tilde{\rho}_a\tilde{u}_b^2},\quad \rho=\frac{\tilde{\rho}-\tilde{\rho}_2}{\tilde{\rho_1}-\tilde{\rho_2}}.
\end{equation}
The Reynolds number $Re$ and the Schmidt number $Sc$ arising in the dimensionless equations \eqref{eq:NS} are defined by:
\begin{equation}\label{eq:prhoadim}
    Re=\frac{\tilde{u}_b\tL_y}{\tilde{\nu}},\quad Sc=\frac{\tilde{\nu}}{\tilde{\kappa}},
\end{equation}
where $\tilde{\nu}$ is the kinematic viscosity and $\tilde{\kappa}$ denotes the molecular diffusivity of the chemical specie producing the density difference. 

\section{Physics Informed Neural Networks}\label{section2}

Artificial Neural Networks (ANNs) are known as universal approximators \cite{Cybenko} so they can in theory learn any continuous function from observed data.
A neural network consists of an input layer, one or more hidden layers, and an output layer. The layers are composed of nodes, also known as neurons, and each neuron is associated with a set of learnable parameters called weights and biases which are used to transform the input signals and compute the output of the model.  
An ANN of $N$  layers performs a series of computations, known as forward propagation, that transform the input vector $x$ through each layer of the network until a final prediction $a^{[N]}$ is obtained. :

\begin{subequations}
\begin{align}
& a^{[1]} = x \in \mathbb{R}^{n_{1}}, \\
& a^{[k]} = \sigma \left( W^{[k]}a^{[k-1]}+b^{[k]} \right) ,~ 2 \leq k < N ,\\
& a^{[N]} =  W^{[N]}a^{[N-1]}+b^{[N]},~ ~k = N ,
\end{align}
\end{subequations} Here $W^{[k]}\in \mathcal{M}_{n_{k}n_{k-1}}(\mathbb{R})$, $b^{[k]} \in \mathbb{R}^{n_{k}}$ are the weights and biases of the $k^{th}$-layers and $\sigma$ is a non linear transformation called activation function. $n_k$ is the number of neurons of the layer $k$.\\
\\ The learnable parameters of the networks : $\boldsymbol{\theta} = (W_{k},b_{k})_{k = 2\dots N}$ are adjusted through the training process until they reach an optimal value $\boldsymbol{\theta^{*}} = (W_{k}^*,b_{k}^*)_{k = 2\dots N}$ that minimizes a loss function $\mathcal{L}$  which penalizes the mismatch between the network predictions and observed data :
\vspace{-0.5cm}
\begin{center}
\begin{align}
\label{argmin}
    \huge \theta^{*}= \underset{\theta}{arg min}  ~\mathcal{L}(\theta),
\end{align}
\end{center} 
most of the time, for regression tasks, the loss function $\mathcal{L}$ is the \emph{Mean Squared Error} between the observations $\{a_{i}\}_{i = 1 \dots N_{\textrm{data}}}$ and the corresponding predictions of the network $\{a^{[N]}(x_{i})\}_{i = 1 \dots N_{\textrm{data}}}$ where $N_\textrm{data}$ is the size of the observed dataset.\\
The weights are updated at each training step, using the gradient descent :
\vspace{-0.5cm}
\begin{center}
\begin{align}
\label{gradient}
    \theta_{n+1} = \theta_{n} - \eta \nabla_{\theta} \mathcal{L}(\theta),
\end{align}
\end{center}

\noindent where $\eta$ is called the learning rate.\\ The key point in the success of neural networks is \textit{backpropagation}. Backpropagation \cite{higham2019deep} is an algorithm for computing the gradients of the loss function with respect to the learnable parameters in a neural network. Nowadays deep learning libraries such as \textit{Tensorflow/Keras} \cite{geron2019hands} and \textit{Pytorch} \cite{pytorch} include automatic differentiation (AD) tools \cite{griewank1989automatic} that allow an efficient computation of these gradients. 
\newline
In the context of \emph{physics-informed neural networks} (PINNs) \cite{raissi2020science}, an ANN $\boldsymbol{v}_{\mathcal{N}}$ is used to approximate the solution $\boldsymbol{v}$ of partial differential equations like :

\begin{subequations}
\label{eq:PINN}
\begin{align}
     &\partial_{t}\boldsymbol{v} + \mathcal{D}_{\boldsymbol{x}}[\boldsymbol{v}]=0,~\boldsymbol{x} \in \Omega,~t\in [0,T], \label{eq:PINN:1}\\
     &\boldsymbol{v}(\boldsymbol{x},0) = h(\boldsymbol{x}),~\boldsymbol{x} \in \Omega , \\
     &\boldsymbol{v}(\boldsymbol{x},t) = g(\boldsymbol{x},t),~ \boldsymbol{x}\in \partial \Omega, t\in [0,T],
\end{align}
\end{subequations} \\
where $\mathcal{D}_{\boldsymbol{x}}$ is a linear or non linear differential operator , $ \Omega,~ \partial \Omega, [0,T]$ are respectively the spatial domain and his frontier, and the temporal domain. The terms $h|_{\Omega}, g|_{\partial\Omega,[0,T]}$ represent initial and boundary conditions, respectively. In this paper, we focus on two-dimensional problems, meaning that the position vector $\boldsymbol{x}$ will be defined in terms of two coordinates, x and y. Thus, $\boldsymbol{x}=(x,y)$.  \\
We define $f$ that represents the left-hand-side of \eqref{eq:PINN:1}:
\begin{center}
    \begin{align}
        f:=\partial_{t}\boldsymbol{v}+ \mathcal{D}_{\boldsymbol{x}}[\boldsymbol{v}],
    \end{align}
\end{center}
$v_{\mathcal{N}}$ takes as input $(\boldsymbol{x},t)$ and AD enables the computation of the derivatives of $\boldsymbol{v}_{\mathcal{N}}$ with respect to $\boldsymbol{x}$ and $t$, thus we can get an explicit expression of $f(\boldsymbol{x},t),~\forall (\boldsymbol{x},t)$.
This stage is the key instance of PINNs because automatic differentiation does not introduce truncation errors due to numerical approximations as with conventional derivation methods.
The objective function to minimize is expressed as :
\begin{center}
\begin{subequations}
\begin{align}
    & \mathcal{L} = \lambda_{1}\mathcal{L}_{f}+\lambda_{2}\mathcal{L}_{b}+\lambda_{3}\mathcal{L}_{0}+\lambda_{4}\mathcal{L}_{\textrm{data}}, \\
    & \mathcal{L}_{f} = \dfrac{1}{N_f}\sum_{i=1}^{N_{f}}||f(\boldsymbol{x}^{f}_{i},t^{f}_{i})||^{2}, \\
    & \mathcal{L}_{b} = \dfrac{1}{N_b}\sum_{i=1}^{N_{b}} ||\boldsymbol{v}_{\mathcal{N}}(\boldsymbol{x}^{b}_{i},t^{b}_{i})-g(\boldsymbol{x}^{b}_{i},t^{b}_{i})||^{2}, \\
    & \mathcal{L}_{0} = \dfrac{1}{N_0}\sum_{i=1}^{N_{0}} ||\boldsymbol{v}_{\mathcal{N}}(\boldsymbol{x}^{0}_{i},0)-h(\boldsymbol{x}^{0}_{i})||^{2}, \\
    & \mathcal{L}_{\textrm{data}} = \dfrac{1}{N_\textrm{data}}\sum_{i=1}^{N_{\textrm{data}}} ||\boldsymbol{v}_{\mathcal{N}}(\boldsymbol{x}^{\textrm{data}}_{i},t^{\textrm{data}}_{i})-\boldsymbol{v}(\boldsymbol{x}^{\textrm{data}}_{i},t^{\textrm{data}}_{i})||^{2}.
\end{align}    
\end{subequations}
\end{center}

Here, $\mathcal{L}_{\textrm{data}}$ is due to additional observations inside the domain, and $\mathcal{L}_{f},~\mathcal{L}_{b},~\mathcal{L}_{0}$ penalize the residual equations, the boundary, and the initial conditions. \\
 The parameters $\lambda_{1\dots 4} \in \mathbb{R}_{+}$ are non trainable weights set manually to manage the ratio between all loss terms. Collocation points for data, residuals and boundary conditions~:~$\{\boldsymbol{x}^{*}_{i},t^{*}_{i} \}_{i=1 \dots N_{*}}$ are often selected on a uniform Cartesian grid. In the machine learning community, ADAM \cite{kingma2014adam}, a first-order gradient-based optimization algorithm for stochastic objective functions, is usualy selected to optimize the parameter $\theta$.\\

\begin{figure}[H]
    \centering 
      \includegraphics[scale = 0.29]{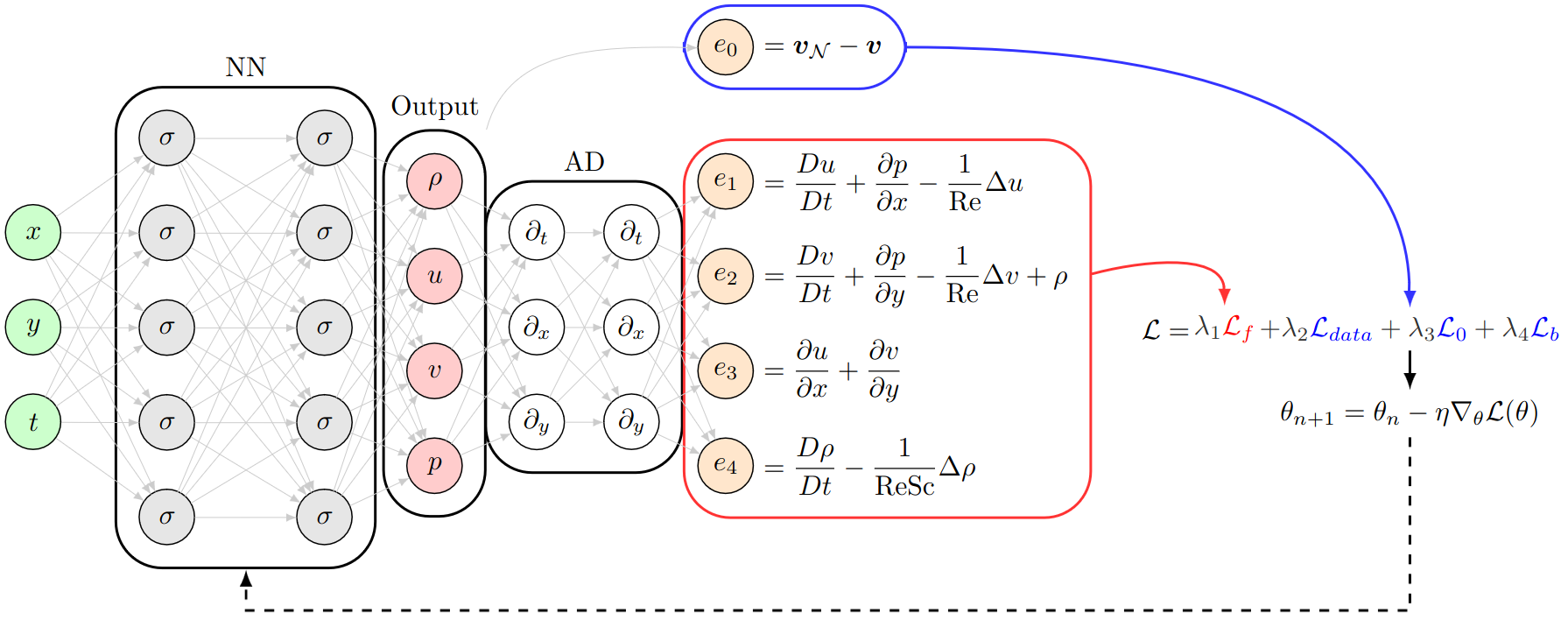}
    \caption{Physic informed neural network structure for gravity currents governing equations: a fully connected neural network take as input $(\boldsymbol{x},t) \in \mathbb{R}^{3}$ and predicts $\boldsymbol{v}_{\mathcal{N}}(\boldsymbol{x},t) = (\rho_{\mathcal{N}},u_{\mathcal{N}},v_{\mathcal{N}},p_{\mathcal{N}})$. The residuals of the governing equations $e_{1,\dots,4}$  are computed by automatic differentiation and $e_0$ denotes the mismatch between the observational data $\boldsymbol{v}$ and the predictions $\boldsymbol{v}_{\mathcal{N}}$, which are combined in the loss function $\mathcal{L}_0, \mathcal{L}_b$ and $\mathcal{L}_\textrm{data}$. Finally, the weights and biases of the network are updated iteratively with gradient descent.  In the residuals, the material derivative expressed as $\dfrac{Df}{Dt}$ inherently includes the convective terms.}
    \label{fig:sketchPINN}
\end{figure}

In the next section, we will apply PINNs to gravity currents generated by lock exchange, to infer the density, velocity and pressure field with synthetical data from 2D simulation. We can see a schematic representation of PINN for this problem Fig. \ref{fig:sketchPINN}.

\definecolor{op}{gray}{0.9}

\section{Application of PINNs on synthetical data}\label{section3}

In this section, we analyze the lock exchange configuration through a single 2D simulation provided by a CFD solver. This approach allows us to access the complete hydrodynamic fields, including density, velocity, and pressure, in both space and time, all derived from just one simulation. Our goal is to train a PINN model using density measurements, treating the simulated density data as if they were experimental LAT data. We reserve the velocity and pressure fields for validation purposes and aim to compare the velocity and pressure fields predicted by the PINN model with the actual fields obtained from the simulation. Evaluating the model in different noise configurations allowed us to find suitable hyperparameters, such as the network structure and activation function. These parameters provide a good compromise between computational time and accuracy and will not be discussed here.
\subsection{synthetic data for the lock-exchange problem}

An accurate representation of the interface between the miscible fluids is crucial in gravity currents simulations, which requires high-order numerical methods to compute steep gradients in the vicinity of the interface. To this end, the governing equations~\eqref{eq:NS} are solved with the code Nek5000 \cite{nek5000-web-page} that has been for example successfully employed in numerical investigations by ~\cite{ozgokmen2004three}. The discretization scheme in Nek5000 is based on the spectral element method proposed by ~\cite{patera1984spectral} with exponential convergence in space and $3^\text{rd}$-order timestepping scheme. 

\begin{figure}
    \centering
    \includegraphics[scale = 0.4]{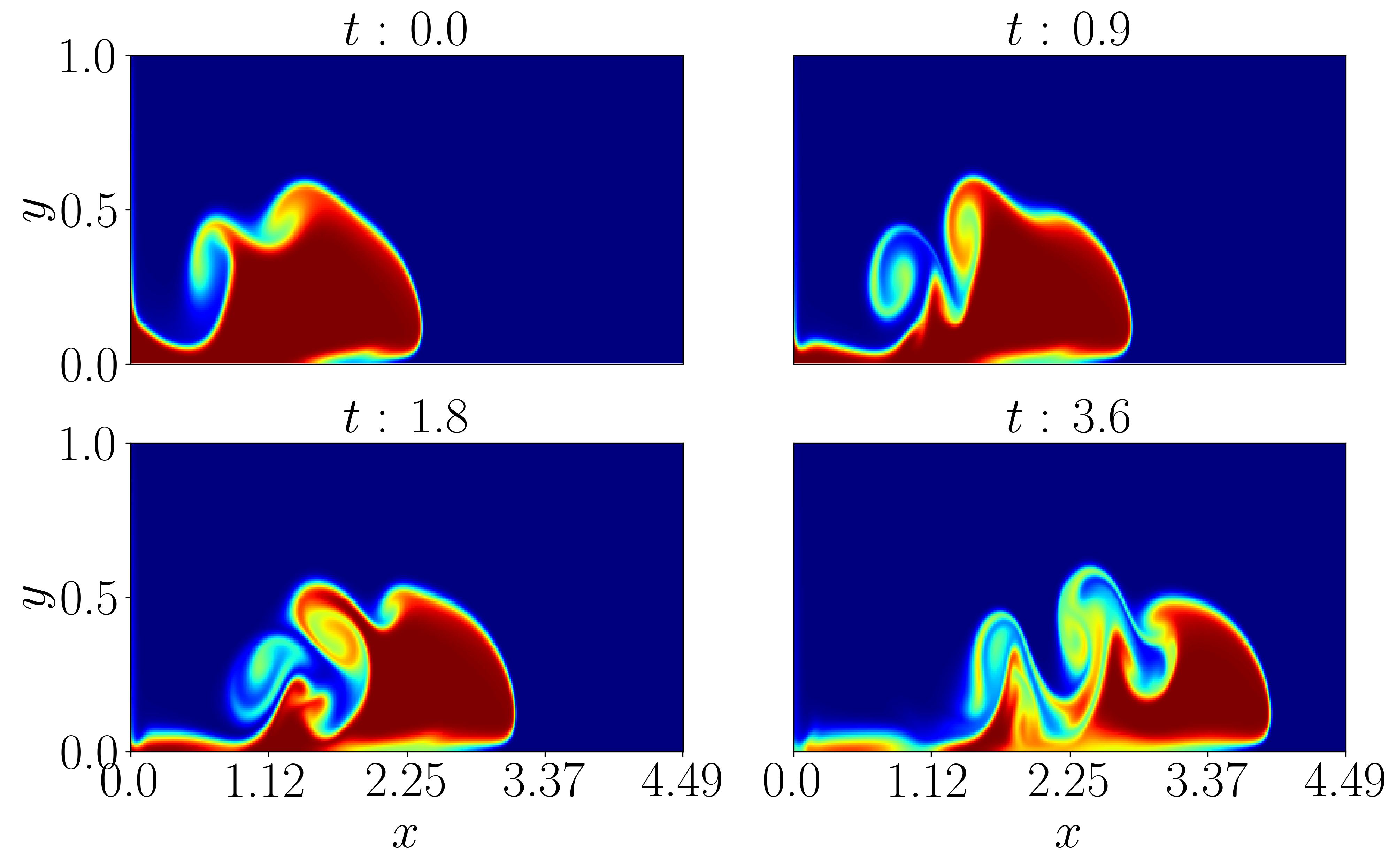}
    \caption{Heatmap of the density field $\rho$  for different time steps during the flow.}
    \label{fig:evolutionrho2D}
\end{figure}

We consider the dimensionless computational domain $\Omega = [0,6.0] \times [0,1.0]$ and $T = [0,10]$ with $Re=5500$. While for liquids such as salt water $Sc\simeq 700$, we consider here $Sc=1$. This assumption is commonly made to reduce computational costs and has low influence on the gravity current front as shown by ~\cite{marshall2021effect}.
\noindent A free-slip condition is applied at $y=1.0$ and no slip-conditions are applied on the remaining domain boundaries. Grid independent results were obtain on a computational mesh of $90\times45$ elements where unknowns are represented by $7^\text{th}$-order Lagrange interpolating polynomials, and with a fixed timestep $\Delta t = 2.10^{-2}$ giving a Courant–Friedrichs–Lewy number less than $0.5$. We show in Fig.~\ref{fig:evolutionrho2D} the time evolution of the density field obtain with Nek5000.

We build from this simulation a training dataset $\mathcal{D}$ where the collocation points are selected in the sub domain: $\Omega_{obs} \times T_{obs}= [0.0, 4.5] \times [0,1.0] \times [4.0,7.5] \subset \Omega \times T $,  which is discretized into a uniform Cartesian grid of $N_x \times N_y \times N_t$ points , where $N_{x}=750$, $N_{y}= 150$ and $N_{t} = 400$.

Finally, to mimic experimental measurements of LAT, we extract from this simulation the observed quantity of the density $\rho$ on each collocation points, and build our training dataset $\mathcal{D} = (\textbf{\textrm{x}}_i,t_i,\rho_i)_{i=1 \dots N_xN_yN_t}$ while the other fields $u,~v,~p$ are used for validation purpose, The dataset $\mathcal{D}$ contains $N = N_xN_yN_t = 45M$ points.

\subsection{Noise robustness}

\begin{figure}
    \centering
    \includegraphics[scale = 0.4]{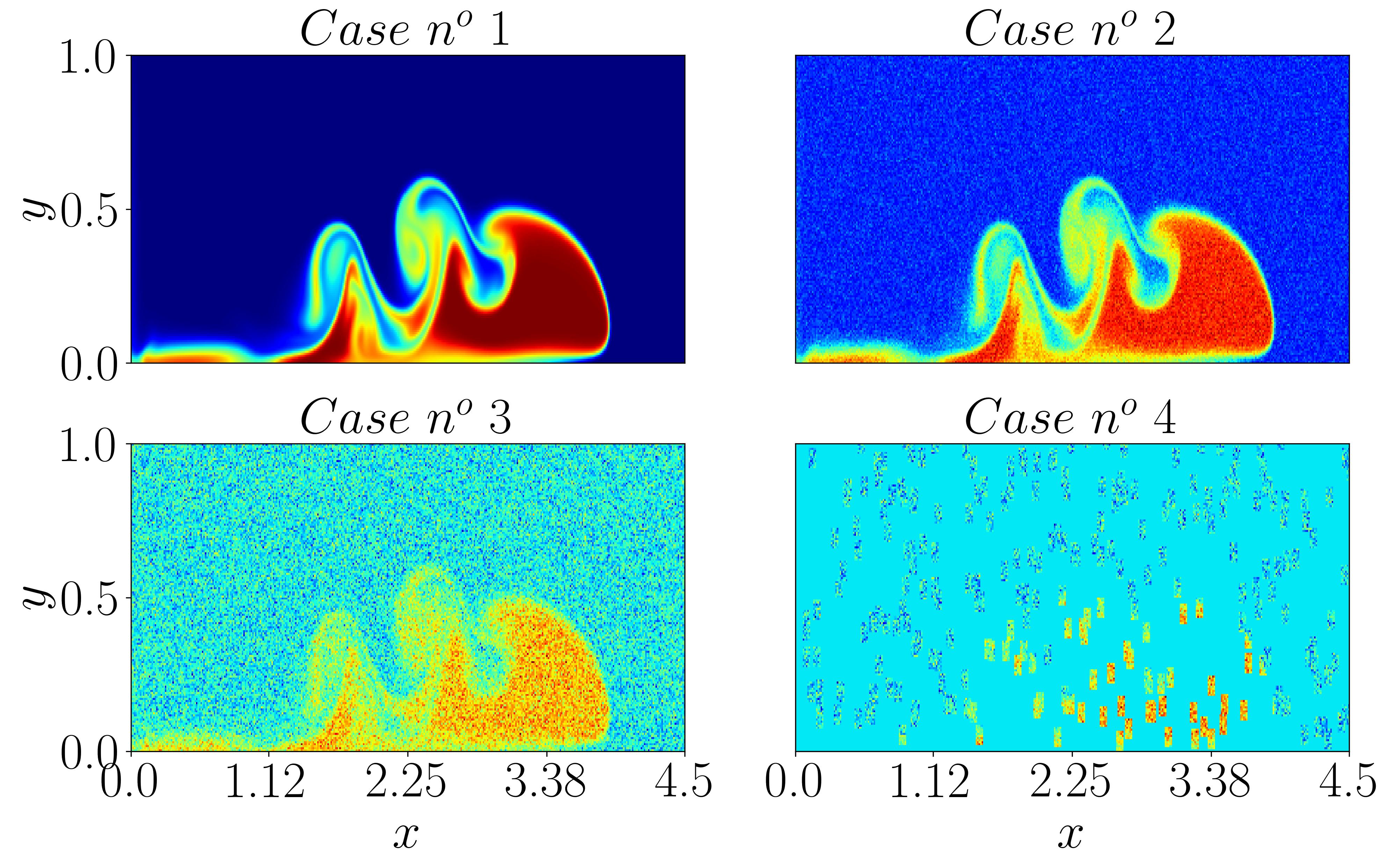}
    \caption{Density field at t = 10 for the different cases studied. In case number 4, 200 squares of size 10 by 10 are randomly selected in the mesh $\Omega_{obs}$}
    \label{fig:vis}
\end{figure}

In this section, noises of varying magnitudes were added to $\mathcal{D}$.
We investigate 4 different cases that are represented in Fig. \ref{fig:vis} : 
\begin{itemize}
    \item \textbf{Case 1: } Original dataset $\mathcal{D}$.
    \item \textbf{Case 2: } We add a gaussian noise to $\rho$ field with standard deviation $\sigma = 5\%$ (experimental precision of LAT).
    \item \textbf{Case 3: } same as Case 2 but with $\sigma =25\%$.
    \item \textbf{Case 4: } Same as Case 3 but in addition, we remove a part of the information in all the snapshots. (we draw 200 squares of $10 \times 10$ pixels randomly from Case 3 dataset). This reduced dataset contains $N = 8M$ points.
\end{itemize}

For all cases, the PINN loss function is expressed as:

\begin{subequations}
\label{eq:Loss2D1}
\begin{align}
    & \mathcal{L} = \lambda\mathcal{L}_{\textrm{data}} + \mathcal{L}_{f},\\
    & \mathcal{L}_{f} = \dfrac{1}{N_f}\sum_{i=1}^{N_{f}}||f(\boldsymbol{x}^{f}_{i},t^{f}_{i})||^{2}, \\
    & \mathcal{L}_{\textrm{data}} = \dfrac{1}{N_{\textrm{data}}} \sum_{i=1}^{N_{\textrm{data}}} ||\rho_{\mathcal{N}}(\boldsymbol{x}^{\textrm{data}}_{i},t^{\textrm{data}}_{i})-\rho(\boldsymbol{x}^{\textrm{data}}_{i},t^{\textrm{data}}_{i})||^{2}. \label{eq:Loss2D1:data}
\end{align}    
\end{subequations}

The neural networks $\mathcal{N}$ take as input $(\textbf{\textrm{x}},t) = (x,y,t)$ and outputs the prediction $\boldsymbol{v}_{\mathcal{N}}(\textbf{\textrm{x}},t) = (\rho_{\mathcal{N}},u_{\mathcal{N}},v_{\mathcal{N}},p_{\mathcal{N}})$. The term $\mathcal{L}_{f}$ penalizes the governing equations ~\eqref{eq:NS} and $\mathcal{L}_{\textrm{data}}$ penalizes mismatch between network prediction for $\rho_{\mathcal{N}}$ and observed $\rho$. The parameter $\lambda$ is a weighting  coefficient used to penalize \eqref{eq:Loss2D1:data}, in practice we set $\lambda = 500$.

\subsection{Results on synthetical data}

The neural network is composed of 8 layers of 250 neurons each. The stochastic Adam optimizer \cite{kingma2014adam} is employed for solving \eqref{argmin}: Glorot normal initializer \cite{glorot2010understanding} is employed to initialize the biases and the weights that are computed iteratively with a gradient update \eqref{gradient} on a subset $\mathcal{B} \subset \mathcal{D}$ called a batch. One training round over $\mathcal{D}$ is called an epoch.  Training is done for 100 epochs with a batch size of 4096. We use an a exponential decay learning rate schedule. The learning rate starts at 1e-3 and ends at 1e-5. Swish  activation function \cite{swishramachandran2017searching} is selected for all layers except the last one.  Finally, training is done with two GPU A6000 in parallel.
\newline

For quantitative results, we define the following relative L2-norm:

\begin{equation}\label{eq:L2gene}
\epsilon_V =\frac{100}{\sup_{\bfx \in  \Omega_{obs} \times T_{obs}} \left| V_\bfx \right|}\sqrt{\frac{\sum_{\bfx\in\bfX}	\left|V_{\mathcal{N}}(\bfx) - V_\bfx\right|^{2}}{\left|\bfX\right|}}
\end{equation}
where $V \in (\rho,u,v,p)$ and $\bfX \subseteq \Omega_{obs} \times T_{obs}  $ defines the domain where the error is computed. We choose $\sup_{\bfx \in \Omega_{obs} \times T_{obs}} \left| V_\bfx \right|$ as reference value to avoid the division near to zero issue reported in \cite{pof}.

\begin{figure}
    \centering
    \includegraphics[scale = 0.45]{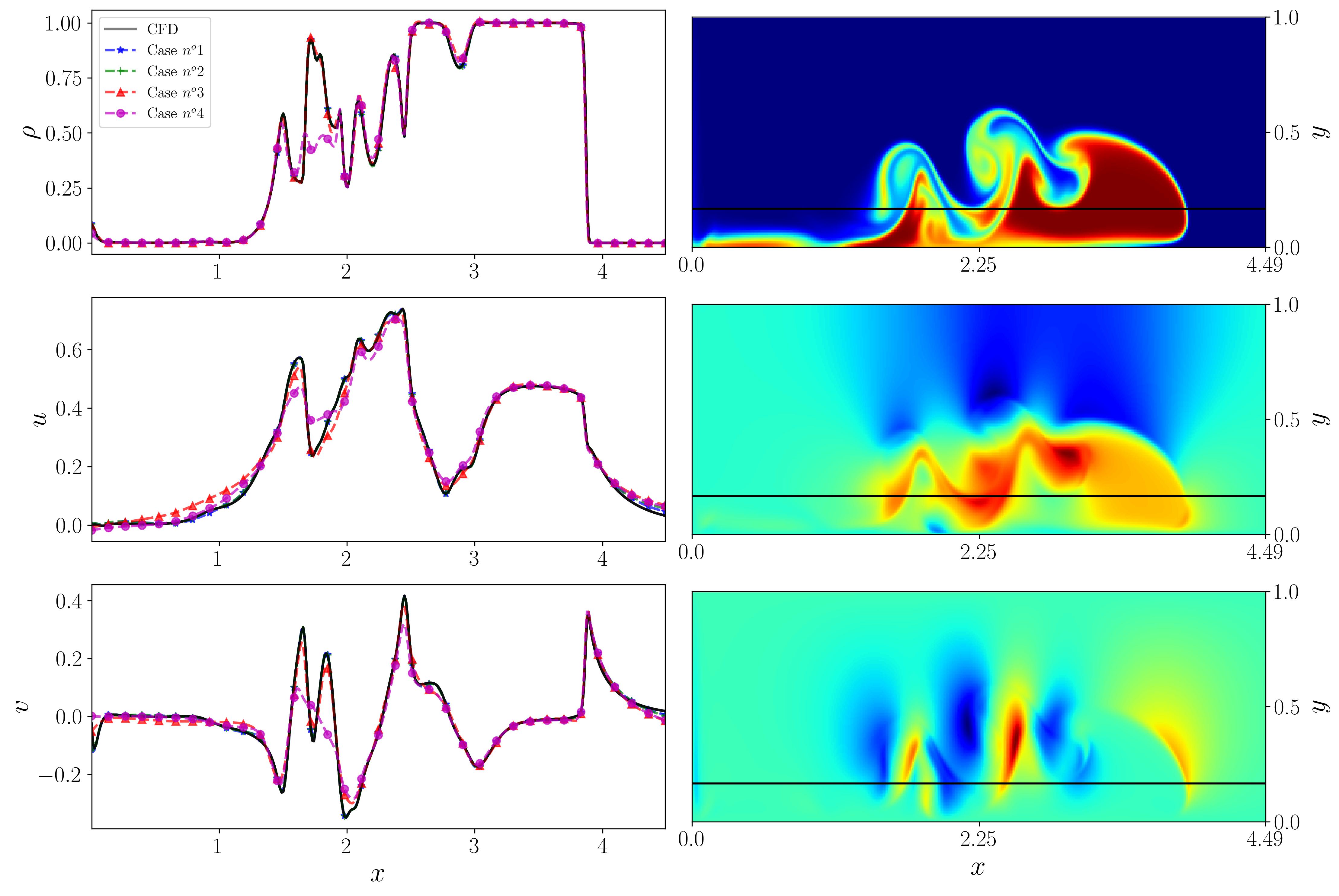}
    \caption{Horizontal profiles comparison. Profiles of $(\rho,u,v)$ reconstructed by PINNs for Cases 1-4 along the x-direction at y = 0.24 and t = 7.5 against the reference solution (left). 
    Corresponding location of the profile in $\Omega_{obs}$ (right).}
    \label{fig:profil1}
\end{figure}

In Fig. \ref{fig:profil1}, the comparison of horizontal profiles shows that the models for Cases 1,2 and 3 are very efficient and can reconstruct all the variations in the current. Moreover, even with 25\% of noise, model for Case 3 is able to filter out measurement noise and correctly reconstruct density and associated velocity fields. The model for Case 4 also performs great mostly everywhere, but we can see a big drop in performance near $x = 1.5$ which may be due to a lack of training data in this area. For all the cases, the results for $u$ and $v$ close to $x = 4.49$ are less accurate, but the steep variations near the front are well inferred.

\begin{figure*}[h]
    \centering
    \includegraphics[scale = 0.35]{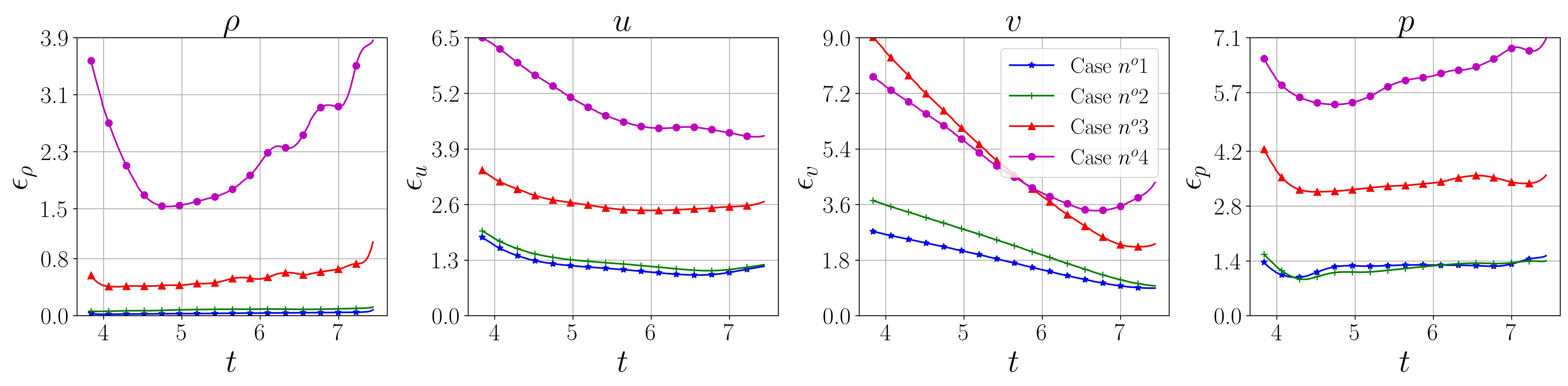}
    \caption{Evolution of the L2-relative error computed on $\boldsymbol{x} \in \Omega$ for density, velocity and pressure fields for $t\in [4, 7.5]$.}
    \label{fig:profil2}
\end{figure*}

Fig. \ref{fig:profil2}  shows the evolution in $T_{obs}$ of the error in $\Omega_{obs}$ for the fields $\rho,~u,~v$. We can see the $L_2$-errors for the density and pressure field are almost constant in time except for density in Case 4. For all cases, the accuracy of the predictions $u$ and $v$ improves with time. Since gravity currents propagate from left to right, at the initial stages of the flow, a large portion of the area covered by the flow contains zero values that might not be useful for training the neural network, making it more challenging to effectively learn and approach the velocity. However, as the gravity current progresses and reaches the end of the flow, the entire current is present in $\Omega_{obs}$, and more useful values such as non zero variations of density can be utilized by the neural network, possibly leading to improved performance.

Table 1 illustrates the global error for Cases 1 to 4 between the density, velocity and pressure fields predicted by PINN, and the corresponding ground thruth. The first three cases have highly accurate predictions with less than 3\% error for all fields, indicating that the model is robust to noise. Despite the higher errors in the fourth case, the prediction errors are still below 8\%, suggesting that the model can handle incomplete data with high noise level.

These results are very promising and the robustness of PINNs to noise in the 2D lock exchange configuration suggests the potential of this framework for real experimental data. In the following section, we apply the PINN model to experimental data obtained by LAT.

\clearpage

\begin{table}[h]
\centering
\begin{tabular}{ p{1.9cm}p{1.1cm}p{1.1cm}p{1.1cm}c  }

Case & $\epsilon_\rho$ & $\epsilon_u$  & $\epsilon_v$  &  $\epsilon_p$  \\
 \hline
Case $n^{o}$1 & 0.035\% & 1.17\%  & 1.81\%  & 1.03\% \\
Case $n^{o}$2 & 0.085\% & 1.29\%  &  2.41\%  & 1.07\% \\
Case $n^{o}$3& 0.53\% & 2.65\% & 5.30\% &2.48\%\\
Case $n^{o}$4 & 7.98\% & 4.93\% & 5.12\% & 4.48\% \\
\end{tabular}
\label{table:numericalL2relative}
\caption{Relative $L_2$-norm errors of the fields reconstructed by PINN computed on the spatio-temporal domain $\Omega_{obs} \times T_{obs}$.}
\end{table}

\section{Application of PINNs to LAT experimental measurements }\label{section4}

\subsection{Experimental setup}
\begin{figure}[H]
  \centering
    \begin{subfigure}{1\textwidth}
    \includegraphics[scale = 0.7]{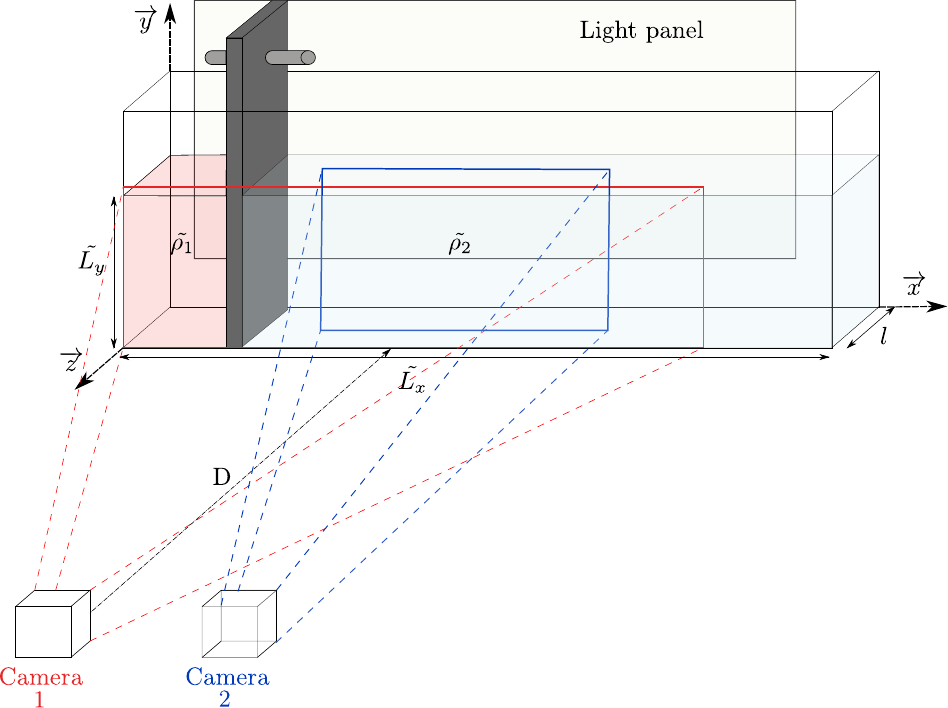}
      \label{fig:aparatusa}
    \end{subfigure}
\begin{subfigure}{1\textwidth}
    \includegraphics[scale = 0.7]{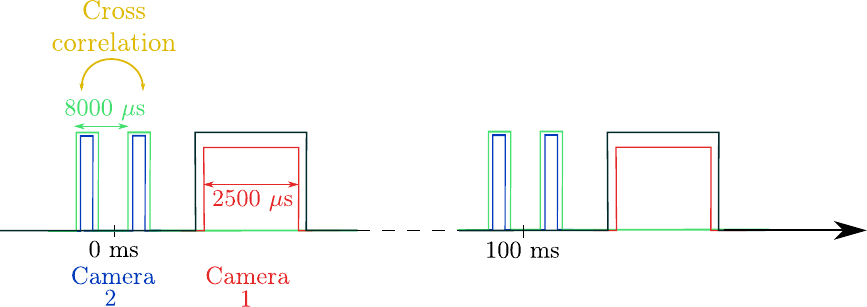} 
      \label{fig:aparatusb}
    \end{subfigure}
    \caption{(a) Sketch of the experimental setup, including the tank,
barrier, light panel, fluids and cameras 1 (LAT) and 2 (PIV). Combination of LAT and PIV measurement devices with $\tilde{L_y}$ = 25 cm,  $\tilde{L_x}$ = 150 cm and l = 10 cm.  The cameras are located at a distance of $D$=250 mm from the tank. (b) Time scale used to represent, in chronological order, the duration of PIV time with 8000 $\mu s$ correlation time between two images and LAT time with 2500 $\mu s$ exposure time.}
\label{fig:apparatus}
  \end{figure}

The experimental apparatus is summarized in Fig.  \ref{fig:apparatus}. Experiments are conducted in a $150$~cm long, $10$~cm wide and $35$~cm high plexiglas tank. After the tank is filled with freshwater to an height $25$ cm, a gate with insulating sides is inserted at 20 cm from the left end of the tank. 

Salt is dissolved in the lock to reach a homogeneous density $\tilde{\rho}_1= 1001.2~\textrm{kg.m}^{-3}$ slightly larger than the freshwater density $\tilde{\rho}_2=999.8~\textrm{kg.m}^{-3}$ which gives us $\Delta \rho = 1.4~\textrm{kg.m}^{-3}$, $Re= 14000$ and $Sc=1000$ according to the definitions in section \eqref{eq:prhoadim}.

Measurements are performed using two cameras and two light sources.
The PIV and LAT cameras are respectively mounted with a $180$ mm f2-8 and $25$ mm lens. 
The cameras are mounted on tripods and placed at a distance of $250$ cm from the tank. Images are recorded at a frame rate of $10$ fps and the spatial resolution of the raw images is 8Mpx for both cameras.

The PIV \cite{raffel2007particle} camera is synchronized with a vertical LASER sheet that illuminates the central section of the tank, providing two-dimensional vertical velocity field measurements as commonly performed in stratified flows.

The LAT camera is synchronized with a white LED panel place behind the tank. The latter technique allows to measure high resolution cross-averaged density fields by adding a known concentration of tracer in the dense compartment. In the present experiment, a red food dye is used as a tracer for salinity and hence density. The reader may refer to \cite{dossmann2016experiments} for a detailed description of the LAT protocol.

The two cameras and light sources are computer-controlled, which permits to synchronize velocity and density measurements. As the LED panel interferes with the PIV measurements, it is required to trigger the PIV and LAT sources in a slightly phase-shifted manner. Measurements are latter recombined on an identical timestep during the PIV post-process. A common spatial origin is used to match the recorded velocity and density field.

As the studied flow evolves in 3D, using the canonical PINN model would imply forcing a neural network to impose the Navier-Stokes equations in 3D and having available measurements of the density field in the whole tank which is prohibitive in our case.

\begin{figure}[H]
    \centering 
    \includegraphics[scale = 0.45]{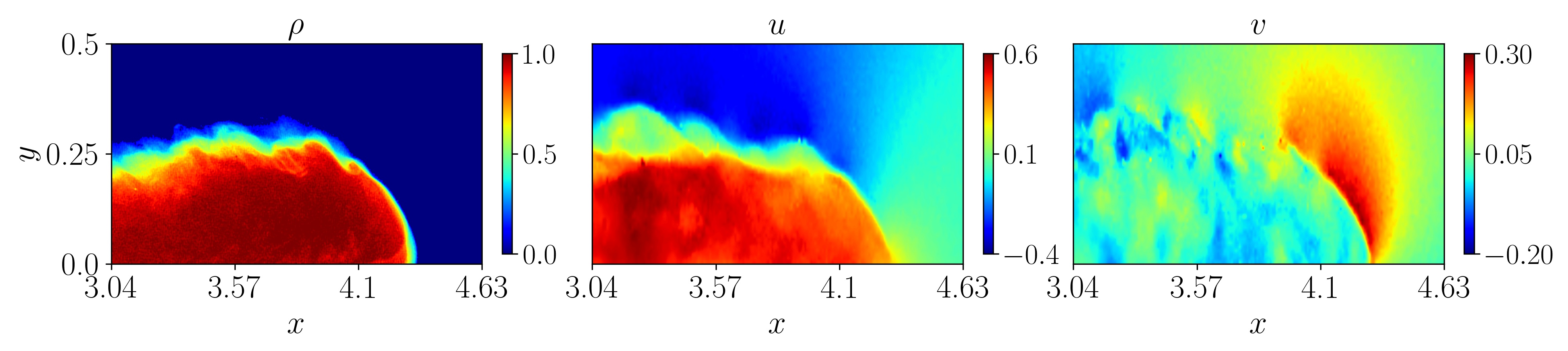}
    \caption{measurements of density at $t = 20.45$ and two components of velocity at $t = 20.80. ~ \rho $ is measured by LAT and $(u, v)$ are measured by PIV in the middle plane of the tank }
    \label{fig:exppivlat}
\end{figure}

So in the following, we assume that the flow is homogeneous in the z-direction, this assumption is commonly made for the lock release configuration \cite{inbook}. As a consequence, we reduce the problem size and consider only the 2D Navier-Stokes equations \eqref{eq:NS} and, under this assumption, the spanwise-averaged density $\overline{\rho}$ measured by LAT is assimilated to 2D density measurements $\rho$.
In Fig. \ref{fig:exppivlat}, we can see a sample of the fields measured by the combination of LAT and PIV. 

\subsection{Training data}

The data are acquired on the spatial and temporal dimensionless domain $\Omega_{exp} \times T_{exp} = [3.0,4.65] \times [0.05,0.5] \times [19.5,21.5]$. This domain is discretized into a uniform grid of $N_x \times N_y \times N_t$ points , where $N_{x}=1050$, $N_{y}= 300$ and $N_{t} = 90$. The training dataset contains $N = N_xN_yN_t = 28350 000$ points of density measurements.

We aim to infer the hydrodynamic field in the middle plane of the domain given LAT measurements of density. We keep the available PIV measurements for validation purposes and do not use the velocity data obtained from PIV to train the model. We apply the canonical PINN method with LAT experimental measurements as observation data in the domain $\Omega_{exp} \times T_{exp}$, considering the governing equations \eqref{eq:NS} and loss function (11). After training the model on LAT density data, we will compare the velocities predicted by the model to the PIV velocities, which will be considered as a test error.

\begin{figure}[H]
    \centering
    \includegraphics[scale = 0.42]{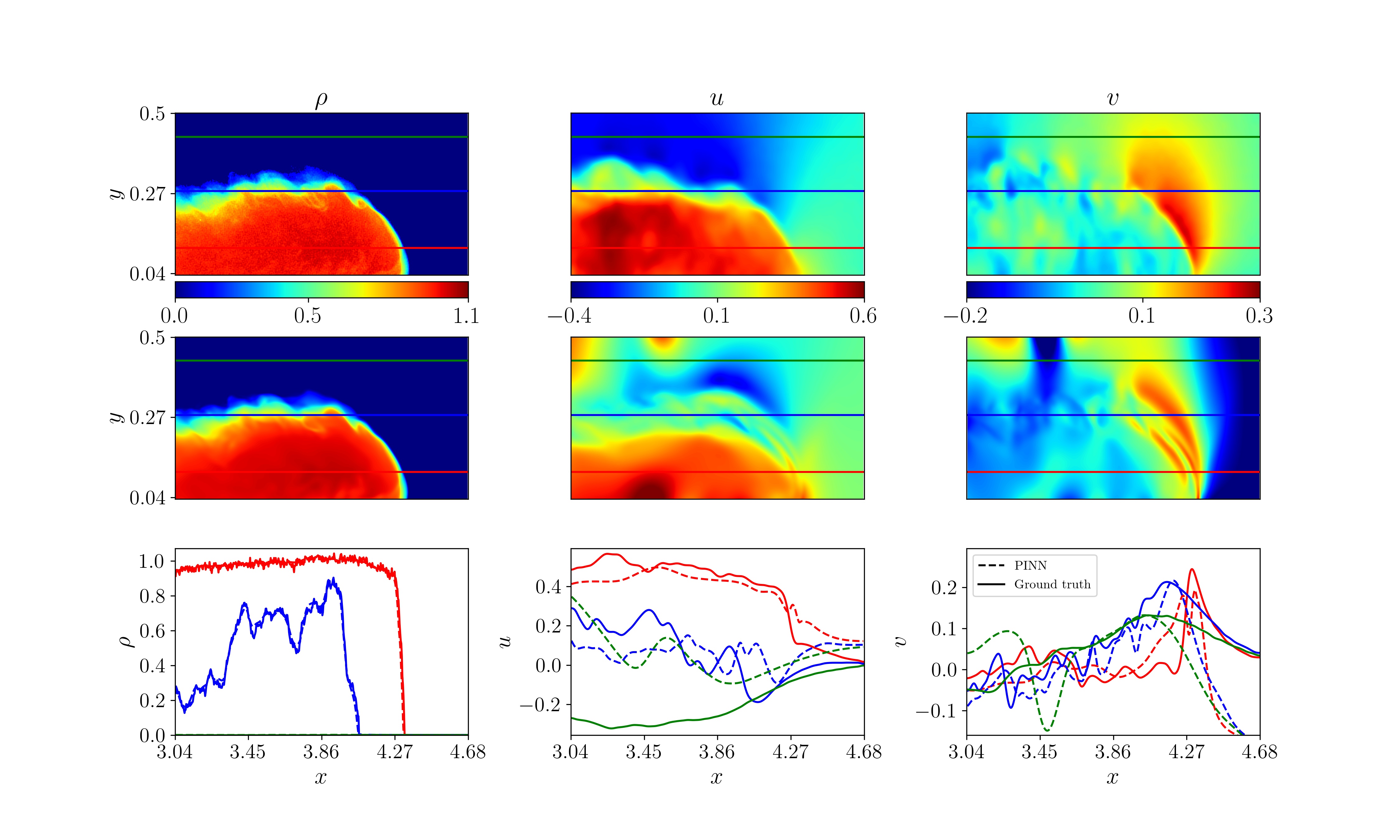}
    \caption{Comparison between measured and predicted density and velocity for model $\mathcal{N}$ at $t=20.8$. First line refers to $\rho$ obtained by LAT and $(u,v)$ measured by PIV considered as ground truth.} Second line refers to corresponding fields inferred by PINN, and third line corresponds to horizontal profiles along x-axis for $y= 0.11$ (red), $y = 0.28$ (blue) and $y = 0.48 $ (green) for both inferred and measured fields.
    \label{fig:profil2div}
\end{figure}

\subsection{Results on experimental data}

The neural network is composed of 8 layers of 250 neurons each which, in our case is the structures which gives the best results. Training is done for 150 epochs with a batch size of 4096. We use Adam optimizer with a decreasing learning rate schedule. The learning rate follows a exponential decay starting from 5e-4 to 1e-5. Finally we use $Swish$ activation function on each layers transition, except for the last layers where no activation function is selected. 

In Fig. \ref{fig:profil2div}, we can see a comparison between predicted $\rho$, $u$ and $v$ fields and the corresponding measured fields by LAT/PIV. Since the model is trained on density, it is not surprising that the predicted density values are highly accurate, but we can notice that PINN has a smoothing effect on the density field. What is the most interesting here is the comparison on the velocity fields. Indeed, the NN has succeeded without prior information on $u$ and $v$ to reproduce the right amplitudes of values as well as the location of the current front. On the other hand, there is a bias in the values for $u$ and the small variations above the current (blue profile) are poorly captured. Moreover, for $v$, the values on the right of the front seem to be outliers.

After a thorough study of the problem parameters, we came to the conclusion that the residual related to the incompressibility equation \eqref{eq:cont} in 2D degraded the results. So, to account for the spanwise variations of the flow, we propose a modification of the continuity equation \eqref{eq:cont} :

\begin{equation}
  \label{eq:cont1}
     \frac{\partial u}{\partial x} + \frac{\partial v}{\partial y} + \xi = 0,
\end{equation}

where $\xi$ depends on $(x,y,t) \in \Omega_{exp} \times T_{exp}$ is a new output of the neural network. 
Thus, the new model $\mathcal{M}$ take as input $(\boldsymbol{x},t) = (x,y,t)$ and outputs the prediction $\boldsymbol{v}_{\mathcal{M}}(\boldsymbol{x},t) = (\rho_{\mathcal{M}},u_{\mathcal{M}},v_{\mathcal{M}},p_{\mathcal{M}},\xi_{\mathcal{M}})$.
Since the 3D Navier-Stokes equations involve $\dfrac{\partial w}{\partial z}$ in the incompressibility equation, we expect $\xi_{\mathcal{M}}$ to converge during training to the restriction of $\dfrac{\partial w}{\partial z}$ to the PIV plane. 
In the following, we denote by $\mathcal{N}$ or $\mathcal{M}$ either if the model was trained with the equation \eqref{eq:cont} or with \eqref{eq:cont1}.

\begin{figure}[H]
    \centering
    \includegraphics[scale = 0.40]{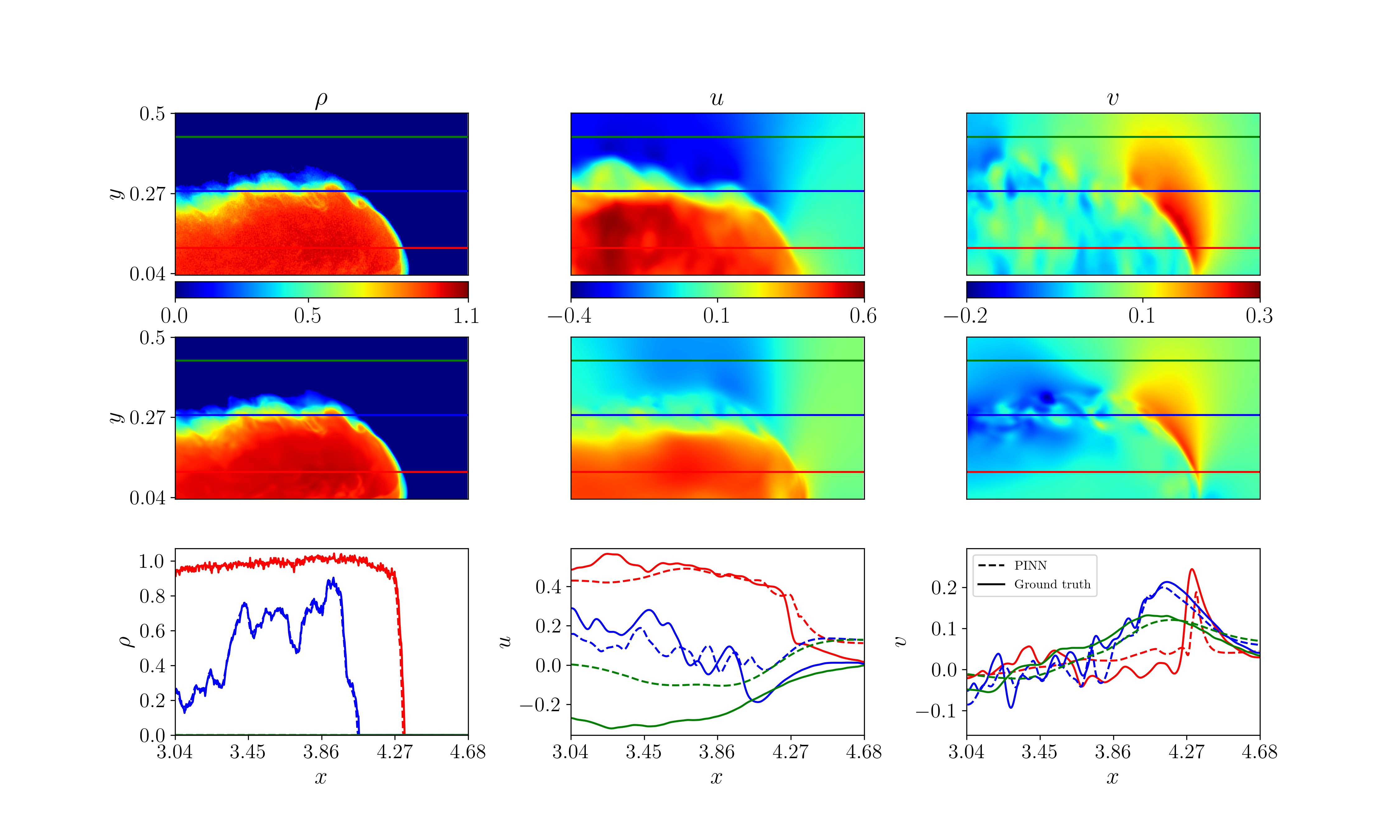}
    \caption{Comparison between measured and predicted density and velocity for model $\mathcal{M}$ at $t=20.8$.  at $t = 20.8$. First line refers to $\rho$ obtained by LAT and $(u,v)$ measured by PIV . Second line refers to corresponding fields inferred by PINN , and third line corresponds to horizontal profiles along x-axis for $y= 0.11$ (red), $y = 0.28$ (blue) and $y = 0.48 $ (green) for both inferred and measured fields.}
    \label{fig:profilw_z}
\end{figure}

The modification made to the equation has led to significant improvements in the accuracy of the predictions.
The results Fig. \ref{fig:profilw_z} show the efficiency of this new formulation of the problem. 
 For the $u$ component of velocity, although a small bias can still be observed on the values above the stream, the predicted quantities inside the stream and near the blue zone are suitable. The predicted  profiles for the $v$ component of velocity in the blue and green zones are now accurate, and the model is able to capture steep variations, such as the large peak at $x=4.3$, with good accuracy. 

\begin{table}[H]
\centering
\begin{tabular}{ p{1.9cm}p{1.1cm}p{1.1cm}c  }
Model & $\epsilon_\rho$ & $\epsilon_u$  & $\epsilon_v$  \\
 \hline
$\mathcal{N}$ & 1.72\% & 20.90\%  &  31.5\%   \\
$\mathcal{M}$   & 1.35\% & 14.6\% & 12.65\% \\
\end{tabular}
\caption{Relative $L_2$-norm errors on the spatio-temporal domain $\Omega_{exp} \times T_{exp}$ between reconstructed and measured fields. model name is $\mathcal{N}$ or $\mathcal{M}$ either if training was done with standard incompressibility equation \eqref{eq:cont} or variation \eqref{eq:cont1}}
\label{table:expL2relative}
\end{table}

The Table \ref{table:expL2relative} shows the overall improvement in predictions brought by the simple change in the{incompressibility equation}  \eqref{eq:cont1}. The model achieve accurate results on the global space $\Omega_{exp} \times T_{exp}$.

\subsection{Quantities Derived from PINNs}

Once trained on experimental density data for a fixed flow, model $\mathcal{M}$ acts as a surrogate model that allows us to calculate any quantity involving density, velocity, or pressure, as well as their derivatives. These values are obtained locally throughout the entire spatial and temporal domain. It would allow us to estimate the vorticity, and energy flux field   $J = p\boldsymbol{u}$ \cite{allshouse2016internal}, which is crucial for understanding the energy budget and fluid circulation in density-varying fluids such as the ocean and the atmosphere. However, it is rarely possible to determine this quantity, which requires simultaneous measurements of the pressure and velocity perturbation fields $p$ and $u$, respectively. Using the PINN model, we present a method for obtaining the instantaneous $J$  from density measurements only, providing a capability that remains out of reach for known methods.

\begin{figure}[H]
    \centering
    \includegraphics[scale = 0.5]{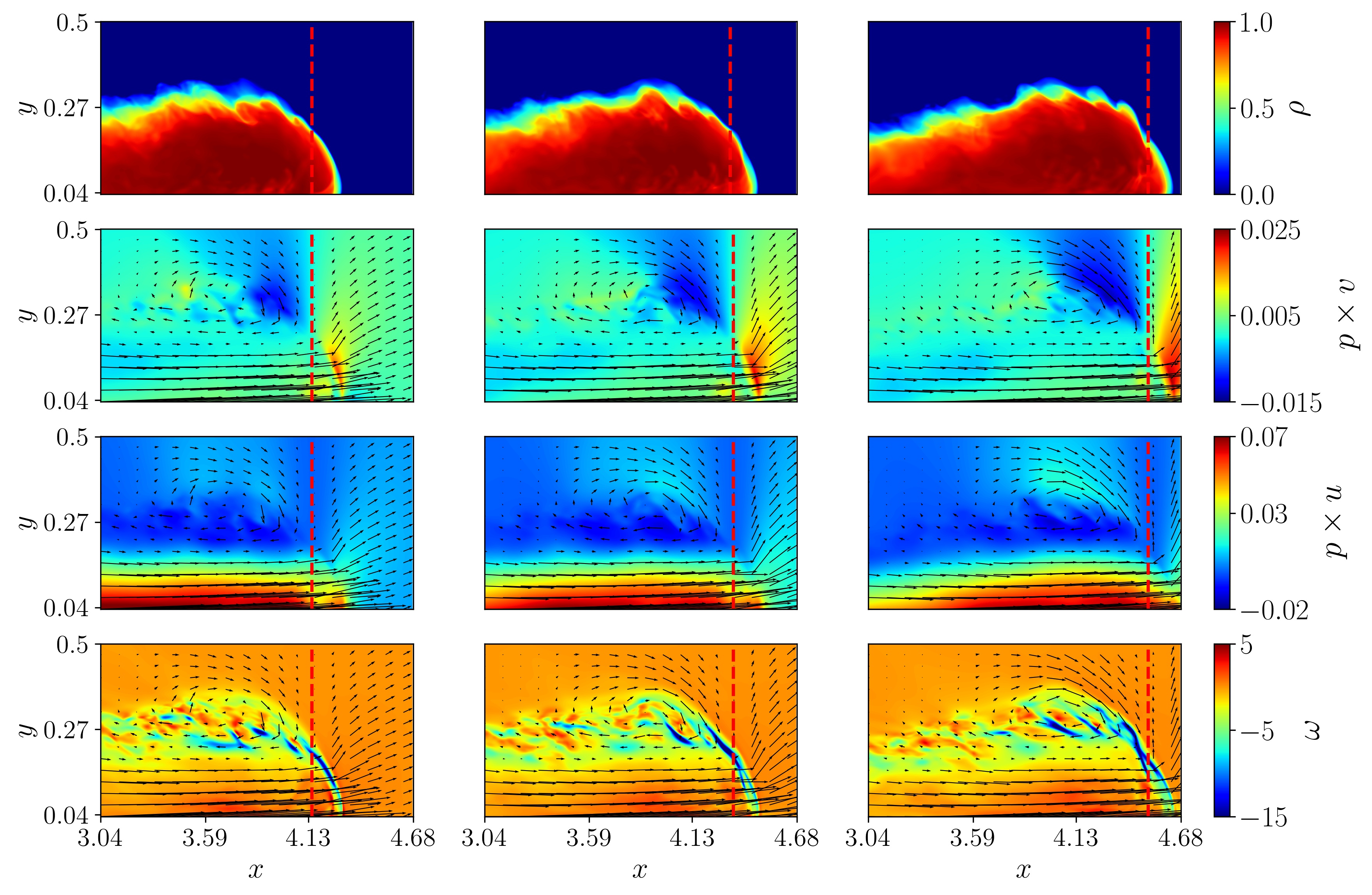}
    \caption{Horizontal and vertical components of the energy flux field $\rho \boldsymbol{v}$, and vorticity. Isovalues of $\rho$ (first line), $pu, pv$ (seconde and third line) and vorticity $\omega$ (fourth line) at times : $t = 20.7, 21.14$ and $21.6$ in addition in lines 2,3,4, the vector field $p\boldsymbol{u}$ is added. }
    \label{fig:stream}
\end{figure}

In Fig. \ref{fig:stream}, we represent the quantity $J = p \boldsymbol{u}$ where $\boldsymbol{u} = (u,v)^{T}$, and respectively the isovalues of the separated components $pv$, $pu$ given by $\mathcal{M}$ at three different times during the flow. These quantities provide information about the energy flux within the gravity current.

We clearly evidence a zone where the current transfers energy to the surrounding fluid, which are the zone where $pv$ reach its maximum values in the right of the current front. In the same way, one identifies a zone where the current captures energy from the ambient fluid which are the zone of circulation above the current where $ pu $ reach its minimum values. We can notice a very sharp transition between these two zones (red dotted line) where $pv$ changes sign. This criterion is consistent through time and represents a quantitative criterium to divide the density field in two parts, the front and the head of the current. In the last raw of Fig. \ref{fig:stream} vorticity isovalues inferred $\omega$ are plotted and superposed to the energy flux vector field. We observe that the transition between the two previously identified zones is also associated with the appearance of re-circulation rolls where the mixing becomes efficient.

\section{Conclusion}

In this work, we quantitatively demonstrate that PINNs are able to solve inverse problems from experimental data. We first investigated the robustness of this model to noisy and gappy data. This study was made from synthetic data obtained by the CFD solver Nek5000 for the lock-exchange configuration, and allowed us to study the parametric aspect of the model. Next, we trained a physics-informed model using experimental density measurements obtained by LAT in the lock-exchange configuration. The inferred velocity fields were then compared to PIV measurements taken simultaneously during the same experiment. Despite the assumption of homogeneity in the z-direction, the results are convincing, and the model succeeds in predicting the pressure field. 
In our previous paper \cite{pof}, we highlighted an optimal experimental setup that would allow the reconstruction of the entire 3D field from multiple 2D measurements of PIV and LAT. In future work, when our experimental setup allows for such measurements, we will consider a case of 3D reconstruction for the lock exchange.

\section*{\label{sec:acknowledgements}Acknowledgements}

A part of this study is conducted in the framework of the Agence Nationale de la Recherche (Project "Pinn'terfaces" ANR-23-CE23-0020 ). 

\section*{Data Availability Statement}
The data that support the findings of this study are available from the corresponding author upon reasonable request.  

\clearpage

\bibliographystyle{unsrt}

\bibliography{biblio}

\end{document}